\documentclass[10pt,showpacs,twocolumn]{revtex4}
\usepackage{graphicx}
\usepackage{amsmath}
\usepackage{amssymb}
\usepackage{}

\begin{document}
\title{Classical simulations including electron correlations for sequential double ionization}
\author{Yueming Zhou$^1$, Cheng Huang$^1$, Qing Liao$^1$ and Peixiang Lu$^{1,2}$ \footnote{Corresponding author: lupeixiang@mail.hust.edu.cn}}

\affiliation{$^1$Wuhan National Laboratory for Optoelectronics and
School of Physics, Huazhong University of Science and Technology,
Wuhan 430074, P. R. China\\ $^2$Key Laboratory of Fundamental
Physical Quantities Measurement of Ministry of Education, China}
\date{\today}

\begin{abstract}
With a classical ensemble model that including electron correlations
during the whole ionization process, we investigated strong-field
sequential double ionization of Ar by elliptically polarized pulses
at the quantitative level. The experimentally observed
intensity-dependent three-band or four-band structures in the ion
momentum distributions are well reproduced with this classical
model. More importantly, the experimentally measured ionization time
of the second electrons [A. N. Pfeiffer {\it et al.}, Nature Phys.
{\bf 7,} 428 (2011)], which can not be predicted by the standard
independent-electron model, is quantitatively reproduced by this
fully classical correlated model. The success of our work encourages
classical description and interpretation of the complex
multi-electron effects in strong field ionization where
nonperturbative quantum approaches are currently not feasible.

\end{abstract}  \maketitle
Among various intense laser-induced phenomena, strong field double
ionization (DI) is one of the most important and fundamental
processes. During the past decades, a great number of experimental
as well as theoretical studies have been performed on this area. It
has been known that DI proceeds either sequentially or
nonsequentially. In nonsequential double ionization (NSDI), the
second electron is ionized by the recollision of the first tunneled
electron \cite{Corkum}. Because of this recollision, the two
electrons from NSDI exhibit a highly correlated behavior
\cite{Weber,Rudenko,Becker,Zhou1,Zhou2}. In sequential double
ionization (SDI), it is usually assumed that no correlation exists
between the two electrons, and thus the ionization of the electrons
can be treated as two independent tunneling-ionization steps.
However, this assumption has been called into doubt by recent
experiments \cite{Fleischer,Pfeiffer}. In Ref. \cite{Fleischer}, it
has been shown that there is a clear angular correlation between the
two electrons from SDI, which implies that the successive ionization
steps are not independent in SDI. In Ref. \cite{Pfeiffer}, the
authors found that the ionization time of the second electron from
SDI is much earlier than the prediction of the independent-electron
model. These observations declare that the electron correlations in
SDI should be reexamined carefully.

Theoretically, an accurate description of the electron correlations
in DI needs full quantum theory. However, due to the enormous
computational demand, the solution of the time-dependent
schr\"{o}dinger equation of the two-electron system in the strong
field, especially in the case of elliptically polarized laser
fields, is not feasible currently. Instead, a fully classical
treatment of the two- and multi-electron systems proposed by Eberly
{\it et al.} has been well established \cite{Ho,Haan,Zhou3,Zhou4}.
During the past decade, this model has been successful in exploring
the strong-field ionization processes at the qualitative level.
However, it fails when made a quantitative comparison with
experiments. For example, in ref. \cite{Wang1} it has been shown
that the saturation intensity for SDI of Ar in the classical
simulation is much higher than the experimental data
\cite{Pfeiffer}, and the amplitude of oscillation in the ratio of
the parallel to antiparallel emitted SDI counts as a function of
intensity is much larger than the experimental observation
\cite{Pfeiffer2}. It remains a question: can a classical treatment
describe the strong field processes at the quantitative level
\cite{Ueda}? In this Letter, we gave an affirmative answer. With a
delicate modification to Eberly's model, we performed a first
quantitative simulation on strong field SDI of Ar in the
close-to-circular laser fields. Our numerical results exhibit
intensity-dependent three-band or four-band structures in the ion
momentum distributions, which are consistent with the recent
experimental results \cite{Pfeiffer}. Especially, the experimentally
measured ionization time of the second electron \cite{Pfeiffer},
which can not be predicted by the standard independent-electron
model, is reproduced surprisingly well with this purely classical
model which fully takes into account the electron correlations
during the entire SDI process. The quantitative agreement between
our numerical results and experimental data indicates that a
classical treatment is a good approximation in describing the
strong-field SDI and has the potential to shed light on the subtle
multi-electron effect in strong-field double and multiple
ionizations.

In the classical picture of a two-electron system, one electron
often drops deeply into the nuclear potential well, leading to the
autoionization of the other electron. In Eberly's classical model
\cite{Ho,Haan,Panfili}, a soft-core potential is introduced for the
ion-electron interaction to avoid autoionization. However, in this
soft-core potential classical model (SPCM), the first and second
ionization potentials can not be matched with those of the
investigated target. In fact, the first electron often ionizes
leaving the second electron with an energy much lower than the
second ionization potential of the target \cite{Bauer,Zhou}. In
tunneling and over-the-barrier ionization, the ionization rate is
very sensitive to the ionization energy. Thus, the SPCM may be
deficient in describing strong field double ionization that occurs
through tunneling or over-the-barrier escape. However, this
deficiency can be avoided with the Heisenberg-core potential, which
not only prevents autoionization but also gives the
ground-configuration energies of the multi-electron atoms
\cite{Kirschbaum}. In the past decades, this potential has been
extensively employed in the classical investigations of atomic and
molecular collisions and laser-matter interactions
\cite{Zajfman,Lerner}. Here, we employ the Heisenberg-core potential
instead of the soft-core potential in the classical two-electron
model to study SDI of Ar by the elliptical laser pulses.

The Hamiltonian of the two-electron atom in the Heisenberg-core
potential classical model (HPCM) is (atomic units are used
throughout this Letter until stated otherwise):
\begin{equation}
H_0=\frac{1}{|{\bf r}_1-{\bf r}_2|}+\sum_{i=1,2}[-\frac{2}{
r_i}+\frac{{\bf p}_i^2}{2}+V_H{(r_i,p_i)}]
\end{equation}
where ${\bf r}_i$ and ${\bf p}_i$ are the position and canonical
momentum of the ${\it i}$th electron, respectively. $V_H{(r_i,p_i)}$
is the Heisenberg-core potential, which is expressed as
\cite{Kirschbaum}:
\begin{equation}
V_H(r_i,p_i)=\frac{\xi ^2}{4\alpha r_i^2}exp\{\alpha[1-(\frac{r_i
p_i}{\xi})^4]\}
\end{equation}
The parameter $\alpha$ indicates the rigidity of the Heisenberg core
and is chosen to be 2. For a given $\alpha$, the parameter $\xi$ is
chosen to match the second ionization potential of the target, i.e.,
it is set to make the minimum of the one-electron Hamiltonian
[$H_1=\frac{-2}{{\bf r}_1}+\frac{{\bf p}_1^2}{2}+V_H{(r_1,p_1)}$]
equal to the second ionization potential of Ar (-1.01 a.u.). For
$\alpha$=2 we obtain $\xi$=1.225. For the two-electron atom, in
refs. \cite{Kirschbaum,Zajfman,Lerner}, the ground-state energy of
the system is obtained by minimizing the Hamiltonian $H_0$. In that
configuration, the two electrons locate at opposite sides of the
nucleus, and they are stationary to each other \cite{Kirschbaum}.
Differently, in our calculation, we employed the approach used by
Eberly {\it et al.} \cite{Panfili} to determine the ground-state
energy of the atom, i.e., by inputting an energy that equals to the
sum of the first and the second ionization potentials of the target
(-1.59 a.u. for Ar). The initial distributions of the ground-state
atom in the phase space are obtained with the approach in SPCM
\cite{Haan,Zhou3}. Figures. 1(a) and 1(b) display the position
distributions of the ground-state atom. We state that in our
calculation, the second ionization potential is related to the
minimum of $H_1$ while the ground-state energy of the atom is
determined by the ``input'' energy, which is higher than the minimal
value of $H_0$. Thus, the two electrons are not fixed at two certain
points but distributed in a finite region of the phase space [see
figs. 1(a) and 1(b)]. This relaxation of the position of the
electron pairs in the phase space allows sufficient electron
correlations in the initial state and during the ionization of the
first electron.

\begin{figure}
\begin{center}
\includegraphics[width=8.0cm,clip]{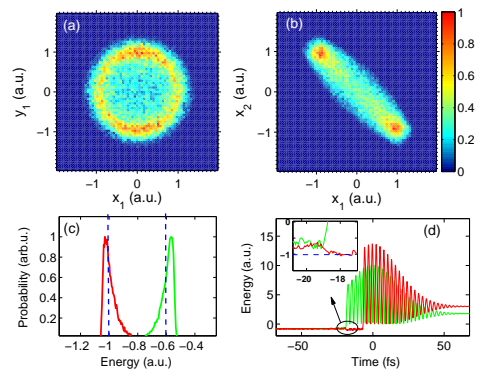}
\caption{\label{fig1} (color online) (a)(b) The initial position
distributions of electrons in the HPCM, where the initial energy of
the system is set to be -1.59 a.u. The parameters of the Heisenberg
core are $\alpha=2$ and $\xi=1.225$. Here ($x_i$, $y_i$, $z_i$)
represent the coordinates of the $i$th electron in the directions of
the $\hat{x}$-, $\hat{y}$-, and $\hat{z}$-axis, respectively. (c)
The ionization energy distributions of the first (the green curve)
and the second (the red curve) electrons. (d) The energy evolution
of the electrons for a typical SDI trajectory.}
\end{center}
\end{figure}

The Hamiltonian of the two-electrons in the presence of the laser
field is
\begin{equation}
H=H_0+({\bf r}_1+{\bf r}_2)\bullet {\bf E}(t)
\end{equation}
where ${\bf E}(t)$ is the electric field of the laser pulses. The
evolution of the system in the laser field is determined by the
following equations:
\begin{equation}
\frac{d{\bf r}_i}{dt}=\frac{\partial H}{\partial {\bf p}_i },
\frac{d{\bf p}_i}{dt}=-\frac{\partial H}{\partial {\bf r}_i }
\end{equation}
The electric field is given as ${\bf
E}(t)=f(t)[\frac{\varepsilon}{\sqrt{\varepsilon ^2+1}}cos(\omega
t+\varphi){\bf \hat{x}}+\frac{1}{\sqrt{\varepsilon ^2+1}}sin(\omega
t+\varphi){\bf \hat{y}}]$, where f(t)=$E_0
exp[-\frac{1}{2}(\frac{t}{\tau})^2]$ is the field envelope.
$\omega$, $\varepsilon$ and $\varphi$ are the laser frequency, the
ellipticity and carrier-envelope phase (CEP), respectively.
$2\sqrt{ln2}\tau$ denotes the pulse duration (FWHM).

First of all, we perform a calculation to test how well the first
and second ionization potentials of Ar have been produced by this
HPCM. The two-electron atoms are exposed to a 33-fs, 788-nm laser
pulse with the intensity I=4.0 PW/cm$^2$ and ellipticity
$\epsilon=0.77$. We trace the energy evolutions of the two electrons
and record the energy of the second electron when the first electron
is ionized. This energy is assumed to be the second ionization
potential of the model atom, which is shown in fig. 1(c). The first
ionization potential of the model atom is also shown in fig. 1(c),
obtained by subtracting the second ionization potential from the
initial energy of the two-electron atom (-1.59 a.u.). It is clearly
shown that the first and second ionization potentials of the model
atom are about -0.57 a.u. and -1.02 a.u. (almost at the bottom of
Hamiltonian $H_1$), respectively, very close to the realistic first
and second ionization potentials of Ar.

\begin{figure}
\begin{center}
\includegraphics[width=7.0cm,clip]{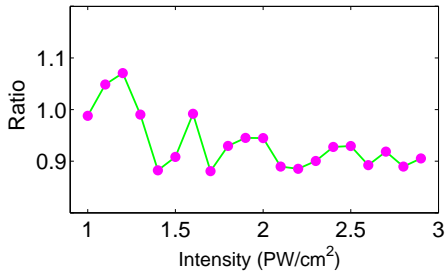}
\caption{\label{fig2} (color online) The ratio of SDI electron
counts of parallel and antiparallel emissions along the minor
elliptical axis ($\hat{x}$ axis) of the laser pulses, as a function
of laser intensity. The laser parameters are the same as those in
ref. \cite{Pfeiffer2}, i.e., the duration, wavelength and
ellipticity of the pulses are 7 fs, 740 nm and 0.78, respectively.
The results are obtained by averaging over the CEP between 0 and
2$\pi$.}
\end{center}
\end{figure}

A recent experiment has shown that in SDI of Ar by strong elliptical
laser pulses the ratio of the parallel and antiparallel electron
emissions along the minor elliptical axis exhibits an oscillating
behavior as a function of laser intensity \cite{Pfeiffer2}. This
behavior has been predicted by the SPCM and explained as a
multi-electron effect beyond independent-electron assumption
\cite{Wang1}. We display our calculations with the HPCM in fig. 2,
where the oscillating behavior is also clear. It implies that the
electron correlations in SDI are included and represented in this
HPCM. One can clearly see from fig. 2 that at the high laser
intensities the oscillating curve is a bit below 1, meaning the two
electrons prefer to emit into the opposite hemispheres. This
behavior well agrees with the experimental data (see fig. 2(a) of
ref. \cite{Pfeiffer2}).

In figs. 3(a) and 3(d) we display the ion momentum distributions in
the polarization plane for SDI of Ar by 33-fs laser pulses with
laser intensities 1.0 PW/cm$^2$ and 4.0 PW/cm$^2$, respectively. The
ellipticity and wavelength are respectively 0.77 and 788 nm, same as
those in the recent experiment \cite{Pfeiffer}. In the direction of
the minor elliptical axis ($\hat{x}$ axis), the distribution
exhibits a three-band structure at the relatively low laser
intensity while a four-band structure at the relatively high laser
intensity. In the direction of the major elliptical axis ($\hat{y}$
axis), as shown in figs. 3(c) and 3(f), the spectra show a gaussian
shape for both intensities. These results agree well with the
experiment \cite{Pfeiffer}. The origin of the four-band structure is
interpreted as due to the different values of the electric field at
which the two electrons are released \cite{Maharjan, Wang2}. Because
of the ellipticity, the electrons preferentially ionize along the
major axis, leading to the electrons with final momenta along the
minor axis. The two outer bands in fig. 3(d) correspond to the
events where the two electrons emit into the parallel directions
whereas the two inner bands result from the events where the two
electrons release into the antiparallel directions. At the
relatively low laser intensity, the momentum amplitudes of the two
electrons are almost the same because both electrons are ionized
around the pulse centre \cite{Pfeiffer}. Thus the antiparallel
electron emissions result in the nearly zero momentum of ion,
leading to the three-band structure in ion momentum distribution
[fig. 3(a)]. The above picture for the intensity-dependent ion
momentum spectra has been confirmed by back analyzing the SDI
trajectories of our calculations (not shown here).

\begin{figure}
\begin{center}
\includegraphics[width=8.0cm,clip]{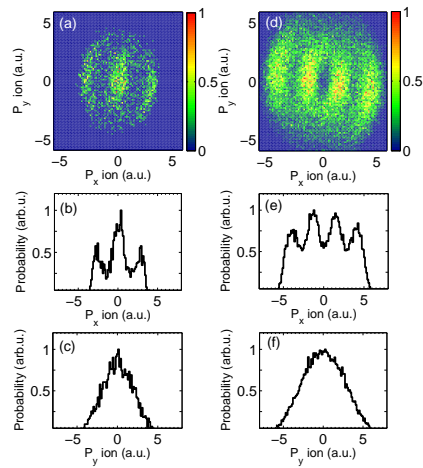}
\caption{\label{fig3} (color online) The ion momentum distributions
in the polarization plane for SDI of Ar at laser intensities of (a)
1.0 PW/cm$^2$ and (d) 4.0 PW/cm$^2$. (b) and (c) show the ion
momentum spectra along the minor and major axis of the polarization,
obtained by integrating the distributions in (a) over $p_y$ and
$p_x$, respectively.(e) and (f) Same as (b) and (c) but for the
distribution in (d).}
\end{center}
\end{figure}

Figure. 4(a) displays the correlated radial momentum distribution of
the electron pairs from SDI of Ar. The laser parameters are the same
as those in fig. 3(d). Note that in this calculation the focal
volume effect assuming a Gaussian beam profile has been considered.
The momentum ${p\\'}_r$ in fig. 4(a) is defined as
${p\\'}_r=\sqrt{[(\epsilon ^2+1)/\epsilon ^2]p_x^2+(\epsilon
^2+1)p_y^2 }$, which is an injective function of time under the
condition that the electron ionized before the peak of the pulse
\cite{Pfeiffer}. The repulsion behavior along the diagonal means
that the two electrons achieve different final momenta at the end of
the pulses, implying the different release times of the two
electrons. In ref. \cite{Pfeiffer}, the ionization times of the two
electrons in SDI are read from the electron's final momentum
${p\\'}_r$. It is found that the ionization time for the first
electron agrees well with the prediction of the independent-electron
model. However, the ionization of the second electrons occurs much
earlier than the prediction of the independent-electron model. With
the same procedure as that in ref. \cite{Pfeiffer}, we extract the
release times of two electrons in our classical simulations where
the electron correlations are included during the entire process of
SDI. The results are shown in fig. 4(b), where the experimental data
from ref. \cite{Pfeiffer} are also displayed. Note that for our
numerical results, the laser intensities in fig. 4(b) are scaled
with a constant factor of 0.82, which is well in the experimental
uncertainty range. Surprisingly, the release times from our
classical simulations agree excellently well with the experimental
data, both for the first and the second electrons. For comparison,
we have repeated the calculations above with the SPCM. The
ionization times of the first and the second electrons from SPCM [as
shown in fig. 4(b)] deviate seriously from the experimental data.
Here, we only displayed the results at intensities above 4.0
PW/cm$^2$ because in the SPCM the DI yield is very low at relatively
low intensities (the DI probability does not exceed 1\% until the
laser intensity reaches 4.0 PW/cm$^2$). During the past decades,
classical methods have been widely employed to describe the strong
field phenomena. However, they are standing at the qualitative point
of view. As far as we know, our calculation is the first successful
example of quantitative investigation of the strong field DI with a
classical treatment.

\begin{figure}
\begin{center}
\includegraphics[width=8.0cm,clip]{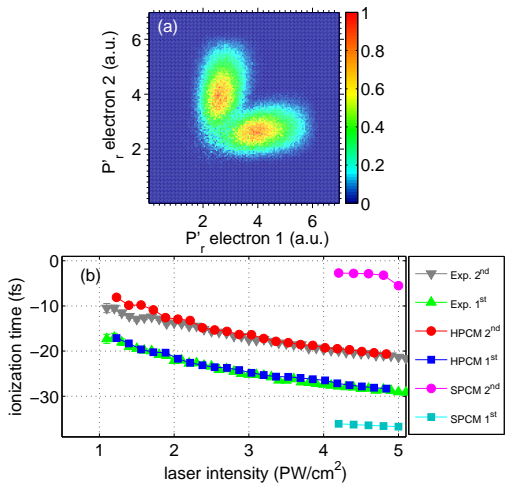}
\caption{\label{fig4} (color online) (a) The electron correlation
spectrum for the momentum ${p\\'}_r$. (b) The release times of the
first (blue squares) and the second (red circles) electrons in SDI,
which are extracted from the electron momentum ${p\\'}_r$ (see text
for detail). The experimental data from ref. \cite{Pfeiffer}
(triangles) are also shown in (b). For comparison, we also displayed
the numerical results from the SPCM (the magenta circles and the
cyan squares).}
\end{center}
\end{figure}

For the theoretical treatment of the independent-electron model in
ref. \cite{Pfeiffer}, the Coulomb interaction between the ion and
the escaping electron is neglected. Here we estimate how strong the
Coulomb interaction affects the release-time reading. It has been
demonstrated that the angular distributions of the electrons can
serve as a signal to estimate the importance of the Coulomb
interaction on the electrons \cite{Popruzhenko,Pfeiffer3}. In fig. 5
we display the angular spectra of the first and the second
electrons. Note that the angular spectra peak at 180$^o$ (and 0$^o$)
if there is no interaction between the ion and the escaping
electron. Fig. 5 shows that the spectrum of the first electron (the
red curves) peaks at an angle slightly deviating from 180$^o$,
indicating a weak Coulomb correction on the electron trajectory
\cite{Popruzhenko,Pfeiffer3}. For the second electron (the green
curves), however, the spectrum peaks almost at 180$^o$, implying
negligible Coulomb interaction between the ion and escaping
electron. These results indicate that the influence of the Coulomb
attraction on the second electron is even weaker than that on the
first electron. Thus, it confirms that the Coulomb attraction has
negligible contribution to the deviation between the experimental
measurement and independent-electron prediction for the ionization
time of the second electron reported in Ref. \cite{Pfeiffer}.

\begin{figure}
\begin{center}
\includegraphics[width=7.0cm,clip]{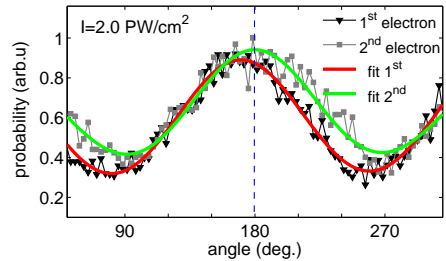}
\caption{\label{fig5} (color online) The angular distributions of
the two electrons from SDI by the 33-fs elliptical
($\varepsilon$=0.77) laser pulses. The solid lines correspond to the
Gaussian fitting of the distributions.}
\end{center}
\end{figure}

It has been proposed that inelastic tunneling \cite{Kornev,Bryan},
where the first electron escapes leaving the core in an excited
state, is important in double and multiple ionization at high laser
intensities. However, the most typical picture of the ionization
process revealed by our classical model is that the first electron
ionizes leaving the second electron at the ground sate of Ar$^+$, as
shown in fig. 1(d) [see the inset of fig. 1(d)]. Note that the
second electron stays at a state with an energy higher than -1.01
a.u. after the first ionization for a small part of the SDI
trajectories shown in figure 1(c). This ``excitation'' is
responsible for the oscillation behavior of the
parallel-antiparallel ratio (see fig. 2) \cite{Wang1}. However, it
is not responsible for the early release time of the second electron
because the second ionization time shown in fig. 4(b) does not
change when these trajectories are excluded. Another reason for the
deviation of the independent-electron prediction of the second
ionization from the measured data \cite{Pfeiffer} possibly ascribes
to the empirical formula for ionization rate \cite{Tong}, which may
be inaccurate in the experimental condition where nonadiabatical
effect is important to the ionization process \cite{Barth}. Then, it
is not clear to what extent the deviation of the second ionization
time comes from the inaccuracy of the ionization rate formula and to
what extent the electron correlations influence the ionization of
the second electron. These questions call for more detailed
investigations.

In conclusion, we have investigated SDI of Ar by the elliptical
laser pulses with the HPCM in which the electron correlation is
included during the entire process. The experimental observed ion
momentum spectra and oscillating behavior of the ratio of the
antiparallel to parallel electron emissions are well reproduced by
this classical model. Especially, the measured ionization time of
the second electron, which strongly deviates from prediction of the
standard independent-electron model, is excellently reproduced by
the HPCM. The quantitative agreement between our classical
calculations and experimental results provides strong supports to
the classical treatment of the multi-electron processes induced by
strong laser fields, which is currently indispensable because the
nonperturbative quantum treatments of the complex effect are not
feasible.

We acknowledge helpful discussions with Dr. A. N. Pfeiffer, Dr. X.
Wang and Prof. J. H. Eberly. We thanks Dr. Pfeiffer for providing us
with their experimental data. This work was supported by the
National Science Fund (No.60925021 and No. 11004070) and the 973
Program of China (No. 2011CB808103).

\end{document}